\documentclass[twocolumn]{article}
\usepackage{graphicx}
\usepackage{txfonts}
\usepackage{psfrag}
\usepackage{natbib}
\def\dsize{\displaystyle}

\begin{document}
   \begin{center}{\bf \Large Stability problem in dynamo }
  \end{center}

%   \subtitle{I. Overviewing the $\kappa$-mechanism}

   \begin{center}{\large M. Reshetnyak}
  \end{center}

   \begin{center}{Institute of the Physics of the Earth, B.Gruzinskaya 10, Moscow, Russia\\
              {\rm email}: {m.reshetnyak@gmail.com} }
  \end{center}

%   \date{Received September 15, 1996; accepted March 16, 1997}

% \abstract{}{}{}{}{} 
% 5 {} token are mandatory
 
\abstract{
  % context heading (optional)
  % {} leave it empty if necessary  
   It is shown, that the  saturated $\alpha$-effect taken from the nonlinear dynamo equations for the
    thin disk  can still produce exponentially growing magnetic field in the case, when this field does not feed back on the $\alpha$. For negative dynamo number (stationary regime) stability is defined by the structure of the spectra of the linear problem for the positive dynamo numbers. Stability condition for the oscillatory solution (positive dynamo number) is also obtained and related to the phase shift of the original magnetic field, which produced saturated $\alpha$ and magnetic field in the kinematic regime.   Results can be used for explanation of the similar effect observed        in the shell models simulations as well in the  3D dynamo models in the plane layer and sphere.
}{ }{ }{ }{ }  
 %  \abstract
  % context heading (optional)
  % {} leave it empty if necessary  
%   {It is shown that the  saturated $\alpha$-effect taken from the nonlinear dynamo equations for the
%    the thin disk  can still produce exponentially growing magnetic field in the case, when this field does not %feed back on the $\alpha$.  It is shown, that it happens, when the linear dynamo problem for the conjugate %problem is stable. We also consider how these results can be used for explanation of the similar effect in the %shell models simulations as well in the  3D dynamo models in the plane layer and sphere.
%  % conclusions heading (optional), leave it empty if necessary 
%   {} {} {} {} {}

  % \keywords{$\alpha\omega$-dynamo, conjugate problem, linear analysis }

  % \maketitle
%
%________________________________________________________________

\section{Introduction}
It is believed, that variety of the magnetic fields observed in astrophysics and   technics can be explained in terms of the dynamo theory, e.g. \citep{HR2004}. The main idea is that kinetic energy of the conductive motions is transformed into the energy of the magnetic field. Magnetic field generation is the threshold phenomenon: it starts when magnetic Reynolds number $\rm R_m$ reaches its critical value $\rm R_m^{\rm cr}$. After that magnetic field grows exponentially up to the moment, when it  already can feed back on the flow. This influence does not come to the simple suppression of the motions and reducing of $\rm R_m$, rather to the change of the spectra of the fields closely connected to constraints caused  by conservation of the magnetic energy and helicity \citep{BS2005}. The other important point is  effects of the phase shift and coherence of the physical fields before and after onset of quenching  discussed in 
 \citep{TB2008}.  
 
 As a result, even after quenching the saturated velocity field is still large enough, so that   $\rm R_m\gg R_m^{\rm cr}$. Moreover, velocity field taken from the nonlinear problem (when the exponential growth of the magnetic field stopped) can still generate exponentially growing magnetic field providing that feed back of the magnetic field on the flow is omitted (kinematic dynamo regime) \citep{CT2009, T2008, TB2008,  SSCH2009}.
 In other words, the problem of  stability  of the full dynamo equations including induction equation, the Navier-Stokes equation with the Lorentz force differs  from the stability problem of the single induction equation with the given saturated velocity field taken from the full dynamo solution: stability of the first problem does not provide stability of the second one. 
 
 Here we consider  effect of such kind of stability  on an example of the model of galactic dynamo in the thin disk, as well as some applications to the  dynamo in the sphere. 
\section{Dynamo in the thin disk}
One of the simplest galactic dynamo models is a one-dimensional model in the thin disk \citep{RSS1988}:
       \begin{equation}\begin{array}{l}\dsize
    {\partial A\over  \partial  t} =\alpha B
 + A''  \\ \\ \dsize 
{\partial B\over  \partial  t} =-{\cal D} A'+ B'', 
\end{array}\label{sys11}
\end{equation}
where   $A$ and $B$ are   azimuthal components of the vector potential  
and magnetic field, $\alpha(z)$ is a kinetic helicity,  
${\cal D}$ is a dynamo number, which is a product of the amplitudes of the $\alpha$- and $\omega$-effects and  primes denote derivatives  with respect to a cylindrical polar coordinate  $z$.
 Equation (\ref{sys11}) is solved in the interval $-1\le z\le 1$ with the boundary conditions 
$B=0$ and $A'=0$ at $z=\pm 1$. We look for a solution of the form
\begin{equation}\dsize
   (A,\,B)=e^{\gamma t} ({\cal A}(z),\, {\cal B}(z)).
\label{sys12} 
\end{equation}
Substituting (\ref{sys12}) in (\ref{sys11}) yields the following eigenvalue problem:
       \begin{equation}\begin{array}{l}\dsize
    \gamma  {\cal A} =\alpha {\cal B}
 + {\cal A}''  \\ \\ \dsize 
\gamma  {\cal B} =-{\cal D} {\cal A}'+ {\cal B}'',
\end{array}\label{sys22}
\end{equation}
where  the constant $\gamma$ is the growth rate. 
 So as   $\alpha(-z)=-\alpha(z)$ is odd function of $z$,  the generation equations   
 have an important property: system (\ref{sys22}) is invariant under transformation $z\to -z$  when \citep{Parker1971}:
\begin{equation}\begin{array}{l}\dsize
{\cal A}(-z)= {\cal A}(z), \quad {\cal B}(-z)= -{\cal B}(z) \\ \\
{\rm or} \\ \\
{\cal A}(-z)= -{\cal A}(z), \quad {\cal B}(-z)= {\cal B}(z).
\end{array}\label{sys33}
\end{equation} 
 Therefore, all solutions may be divided into two groups:  odd on ${\cal B}(z)$, dipole ($\rm D$), and even, quadrupole on ${\cal B}(z)$. Then we can replace  $-1\le z\le 1$ with  the interval $0\le z\le 1$ and the following boundary conditions at $z=0$: ${\cal A}'=0$, ${\cal B}=0$ (D) and ${\cal A}=0$, ${\cal B}'=0$ ($\rm Q$). Usually, $\alpha=\alpha_0$ with $\alpha_0(z)=\sin(\pi z)$  is used, see also \citep{Soward1978} for $\alpha_0(z)=z$ dependence, more  appropriate  for  analytical applications.
 
 System  (\ref{sys22}) has growing solution, $\Re\gamma>0$, when $|{\cal D}|>|{\cal D}^{\rm cr}|$.  
 For  ${\cal D}<0$ the first exciting mode  is quadrupole with ${\cal D}^{\rm cr}\approx -8$ and  $\Im\gamma =0$: solution is non-oscillatory\footnote{For our Galaxy usual estimate is  ${\cal D}=-10$.}. For  ${\cal D}>0$ the leading mode is oscillatory dipole, $\Im\gamma \ne 0$ with 
  higher threshold of generation: ${\cal D}^{\rm cr}\sim 200$.  
 Putting nonlinearity of the form
 \begin{equation}\begin{array}{l}\dsize
 \dsize \alpha(z)={\alpha_0(z)\over 1+E_m} 
 \quad {\rm for}\quad |{ B}|\gg 1
\end{array}\label{non}
\end{equation} 
in  (\ref{sys11}), where  $\dsize E_m=( B^2+{ A'}^2)/2$ is a magnetic energy, gives stationary solutions for $\rm Q$-kind of symmetry and quasi-stationary solutions for $\rm D$, see about various forms of nonlinearities in \citep{Beck}. The property of the nonlinear solution is mostly defined by the form of the first  eigenfunction.
 
\begin{figure}
   \centering
\psfrag{t}{$t$}
\psfrag{M}{$t_0$}   
\psfrag{h}{${\cal C}=1,\, {\cal C}=10$}
\psfrag{n}{${\cal C}=0$}
\psfrag{EAS}{$E_m,\, \widehat{E_m}$}
 \hskip -2cm \includegraphics[width=9cm]{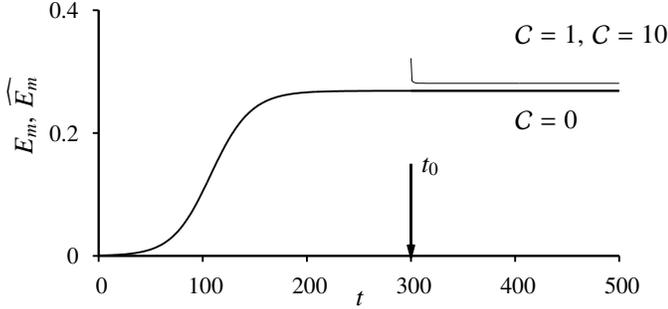}
      \caption{Evolution of magnetic energy $E_m$ for $t<t_0$ governed by the system (\ref{sys11},\ref{non}) for ${\cal D}=-10$. In the moment $t_0=300$ the new magnetic field $(\widehat{A},\, \widehat{B})$  with the initial conditions defined by constant $\cal C$ is switched on, see (\ref{sys44},\ref{non}). Plots for $E_m$ and  
      $\widehat{E_m}$ (with ${\cal C}=0$) for $t>t_0$ coincide. All the solutions are stationary. 
              }
         \label{Fig1}
   \end{figure}

Now,  in the spirit of  \citep{CT2009, TB2008} we add to   (\ref{sys11}) equations for the new magnetic field  $(\widehat{A},\, \widehat{B})$ with the same   $\alpha$  (\ref{non}), which  depends on $({A},\, {B})$ and does not depend on  $(\widehat{A},\, \widehat{B})$:
       \begin{equation}\begin{array}{l}
       \dsize
    {\partial A\over  \partial  t} =\alpha {B}
 + A''  \\ \\ \dsize 
{\partial B\over  \partial  t} =-{\cal D} A'+ {B}'' \\ \\
      \dsize
    {\partial \widehat{A}\over  \partial  t} =\alpha \widehat{B}
 + \widehat{A}''  \\ \\ \dsize 
{\partial \widehat{B}\over  \partial  t} =-{\cal D} \widehat{A}'+ \widehat{B}''. 
\end{array}\label{sys44}
\end{equation}
Numerical simulations  demonstrate that for the negative ${\cal D}$ the both 
  $(A,\, B)$ and   $(\widehat{A},\, \widehat{B})$ are steady, however the final magnitudes of 
  $(\widehat{A},\, \widehat{B})$ depend on the initial conditions for $(\widehat{A},\, \widehat{B})$, see
   Fig.~\ref{Fig1}. The procedure was the following: equations (\ref{sys11}, \ref{non}) for $(A,\, B)$ were integrated up to the moment  $t=t_0$, then the full system  (\ref{sys44}) was simulated with 
   the initial conditions for $(\widehat{A},\, \widehat{B})$ in the form: 
   $\dsize(\widehat{A},\, \widehat{B})\Big{|}_{t=t_0}=(A,\, B)\Big{|}_{t=t_0}(1+{\cal C}\varepsilon)$, where   $\varepsilon \in [-0.5,\, 0.5]$ is a random variable and $\cal C$ is a constant. The both vectors  
   $(A,\, B)$ and   $(\widehat{A},\, \widehat{B})$  are stable in time, however the final magnitude  of
      $\widehat{  E_m}$  for ${\cal C}\ne 0$ slightly depends on  ${\cal C}$.
 Presence of  alignment of the fields        
$(A,\, B)$ and   $(\widehat{A},\, \widehat{B})$ follows from linearity and homogeneity  of equations for 
  $(\widehat{A},\, \widehat{B})$,  where $\alpha(z,\, E_m)$ is given. Latter we consider 
    stability of $(\widehat{A},\, \widehat{B})$  in more details. 
   
\begin{figure}
%   \centering
\psfrag{t}{$t$}
\psfrag{M}{$t_0$}   
\psfrag{h}{${\cal C}=0$}
\psfrag{n}{${\cal C}=10^{-2}$}
\psfrag{EAS}{$E_m,\, \widehat{E_m}$}
 \hskip -1cm \includegraphics[width=9cm]{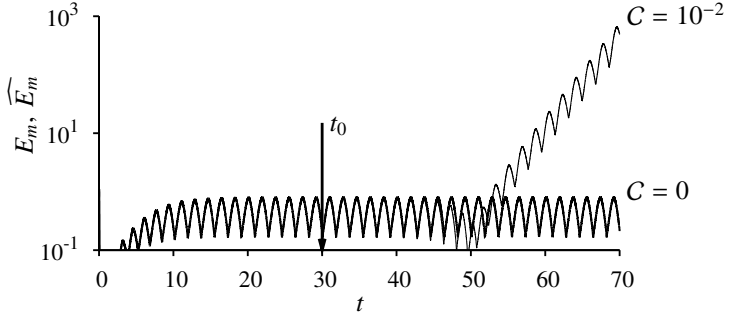}
      \caption{Evolution of magnetic energy $E_m$ for $t<t_0$ governed by the system (\ref{sys11},\ref{non})  for ${\cal D}=300$. In the moment $t_0=30$ the new magnetic field $(\widehat{A},\, \widehat{B})$  with the initial conditions defined by constant $\cal C$ is switched on, see (\ref{sys44},\ref{non}). Plots for $E_m$ and  
      $\widehat{E_m}$ (with ${\cal C}=0$) for $t>t_0$ coincide. For ${\cal C}\ne 0$ after intermediate regime the phase shift $\theta$ between $E_m$ and  
      $\widehat{E_m}$ increased  and exponential growth of  $\widehat{E_m}$ started.   }
         \label{Fig2}
   \end{figure}

 For ${\cal D}>0$ situation is different, resembling that one of instability described in \citep{CT2009, T2008, TB2008,  SSCH2009} for more sophisticated  models: field  $(\widehat{A},\, \widehat{B})$ oscillates and  starts to grow exponentially, 
see  Fig.~\ref{Fig2}. Note, that no regime in oscillations for $(\widehat{A},\, \widehat{B})$ is observed. The other specific feature is delay of $(\widehat{A},\, \widehat{B})$ relative to $(A,\, B)$: $\dsize\theta\approx-{\pi\over 3}$.

 If $E_m$ in (\ref{non}) is averaged over the space, so that $\alpha$ is steady, then instability dissapears. The question arises: does instability depends on stationarity, either it depends on something else?

It is known, that for ${\cal D}<0$  stability of the system  (\ref{sys22}, \ref{non}), which has stationary solution, is tightly bound to behaviour  of the linear solution of (\ref{sys22}) for  ${\cal D}>0$ \citep{RSS1992}. Note, that for the complex form of (\ref{sys22})  it is equivalent  to the solution of the conjugate problem.

Let  $(\widetilde{A},\, \widetilde{B})=({\cal A}+{ a},\, {\cal B}+{ b})$, where  $({\cal A},\, {\cal B})$ is a solution of the nonlinear problem and  $({ a},\, { b})$ is a perturbation with the same boundary conditions as for $({\cal A},\, {\cal B})$. Putting  $(\widetilde{A},\, \widetilde{B})$ in  (\ref{sys22}) with  $\dsize \alpha\approx\alpha_0+{\partial\alpha\over\partial {\cal B}} { b}$  
 yields equations for  $({ a},\, { b})$\footnote{It  is usually  supposed that in $\alpha\omega$-dynamo models $B\gg A'$. }:
        \begin{equation}\begin{array}{l}\dsize 
    \gamma  a =\alpha^{\rm e} b + a''  \\ \\ \dsize 
\gamma  b =-{\cal D} a'+ b'',
\end{array}\label{sys333}
\end{equation}
where $\dsize\alpha^{\rm e}=\alpha+{\partial \alpha\over\partial {\cal B}}{\cal B}$ for 
$\dsize\alpha={\alpha_0\over 1 + {\cal B}^2}$  is 
\begin{equation}\dsize
  \alpha^{\rm e}={1-{\cal B}^2\over \left(1+{\cal B}^2\right)^2}\alpha_0\sim -{\alpha_0\over {\cal B}^2}\quad {\rm for}\quad |{\cal B}|\gg 1.
\label{non1}
\end{equation} 
Behaviour of $\alpha\omega$-dynamo   (\ref{sys22}) is defined by the sign of  ${\cal D}\alpha$, and its change 
 in the perturbed equations  (\ref{sys333}) is important. In other words, instead of nonlinear equations 
   (\ref{sys22},\ref{non}) we come  to the linear problem  (\ref{sys22}) with given $\alpha=\alpha(z,\, E_m)$ and effective dynamo number $\dsize {\cal D}^{\rm e}= -{{\cal D}\over {\cal B}^2}$. 
   Then stability of fields  $(\widehat{A},\, \widehat{B})$ for the  negative 
      $\cal D$ can be explained as follows. For negative  ${\cal D}$ solution $(\widehat{A},\, \widehat{B})$ is finite and stable, because the threshold of generation ${\cal D}^{\rm cr}_{+}$  for  (\ref{sys22})  is much larger than   $\dsize {\cal D}^{\rm e}$, ${\cal D}^{\rm cr}_{+}\gg {\cal D}^{\rm e}$. Field $(\widehat{A},\, \widehat{B})$ is defined up to an arbitrary factor, what  corresponds to  alignment of the vectors  $(A,\, B)$  and $(\widehat{A},\, \widehat{B})$. Note, that ${\cal D}^{\rm cr}_{+}\ll {\cal D}^{\rm e}$ does not guarantee, that $(\widehat{A},\, \widehat{B})$ will grow exponentially due to nonlinearity  (\ref{non}). 
   
It is worthy of note that nonlinear solution of  (\ref{sys11},\ref{non}) and (\ref{sys44},\ref{non}) demonstrates   similar stationary 
 behaviour even  for ${\cal D}\sim -10^{3}$ in spite of the fact, that  ${\cal D}^{\rm cr}_{+}$ 
 for the quadrupole oscillatory  mode for positive ${\cal D}$ 
   is  $\sim 200$. The reason is that dynamo system tends to the state of the strong magnetic filed with 
   ${\cal B}\sim {\cal D}^{1/2}$, so that $\dsize \alpha\sim {1\over {\cal B}^2 }$, leaving ${\cal D}^{\rm e}$ at the   level of the first mode's threshold of generation.       
      
      For positive  
 $\cal D$ $(A,\, B)$, and therefore $\alpha(B)$,  oscillate and one needs additional information 
 on correlation of the waves.    
  Here, instead of (\ref{non1})  we get 
$\dsize \alpha^{\rm e}\sim -{\alpha_0 \widehat{B}\over |B|^3}$. If phase shift between $B$ and $\widehat{B}$ is negligible, then $\alpha$-effect is saturated and time evolution of $(A,\, B)$ and $(\widehat{A},\, \widehat{B})$ is similar. However, simulations demonstrate Fig.~\ref{Fig2}, that  field $(\widehat{A},\, \widehat{B})$ delays relative to $(A,\, B)$.
   This is typical situation, when parameter resonance takes place:  $\alpha$ is modulated by signal with frequency $\Omega\sim 2\omega$, $\omega=\Im\gamma$, see \citep{DP2008} for details of spatial resonance. This assumption is supported by the fact, that instability disappears when in (\ref{non}) steady $\alpha$, averaged in time, is used. 
 Note, that usage of quadrupole boundary conditions in  (\ref{sys44}) is not important for instability: problem 
  (\ref{sys44},\ref{non}) with periodical boundary conditions has oscillatory solution and instability
   depends on the form of quenching in the  same way.  
 
 To demonstrate what happens we consider how delay  $\theta$ of $(\widehat{A},\, \widehat{B})$ relative to
   $(A,\,B)$  
  changes  production of $\widehat{A}^2+\widehat{B}^2$ near the  threshold of generation ${\cal D}^{\rm cr}_{+}$. We start from the  linear analysis of the system in the form:   
       \begin{equation}\begin{array}{l}
       \dsize
   i\omega  \widehat{A} =\alpha \widehat{B}
 -k^2 \widehat{A}  \\ \\ \dsize 
 i\omega  \widehat{B} =-i{\cal D}^{\rm cr}_{+} k \widehat{A}-k^2 \widehat{B}.
\end{array}\label{sys44aa}
\end{equation}  
  From condition of solvability for (\ref{sys44aa}): $\dsize(k^2+i\omega)^2= -i{\cal D}^{\rm cr}_{+} k\alpha_0$  with $\alpha=\alpha_0$ follows that $\omega^2=k^4=1$. The other prediction of the linear analysis is the phase shift $\varphi$ between 
  $\widehat{A}$ and $\widehat{B}$:   $\dsize \varphi=\pm {\pi\over 4}$, what is twice smaller than for the nonlinear regime \citep{TB2008}, so that for the nonlinear regime the  maximal $\widehat{A}$ is when $\widehat{B}$ is zero and  quenching is absent. 
   
Then putting  in (\ref{sys44}) 
$B=b\sin(x-t)$, $\widehat{A}=\sin(x-t+\varphi+\theta)$, $ \widehat{B}= \sin(x-t+\theta)$, and 
 $\dsize\alpha={1\over 1+B^2}$ we get how generation depends on $\theta$. 
   Equation for $\widehat{B}$ does not include original field $(A,\,B)$, so we consider only production of $\widehat{A}^2$. Then 
  $\delta \widehat{A}(\varphi,\,\theta)=\alpha_0\int\limits_0^{2\pi}{\dsize \widehat{B}\widehat{A}\over\dsize 1+B^2}\, dt$. 
  If 
  $\dsize|\Pi|\gg 1$, where $\dsize\Pi={\delta \widehat{A}(\varphi,\,\theta)\over \delta \widehat{A}(\varphi,\,0)}$, then $(\widehat{A},\, \widehat{B})$ is  unstable.
  
   The exact equation for $\delta \widehat{A}$ is:
       \begin{equation}\begin{array}{l}
       \dsize 
\delta \widehat{A}(\varphi,\,\theta)=h_1  
  +
 h_2  \tan(\varphi),
\\ \\
\dsize 
h_1={
  \cos(\theta)^2(4
-3 2^{1/2}) -2(2^{1/2}-1)
\over  2^{1/2}(2^{1/2}-1)},\\  \\ \dsize
h_2={\sin(2\theta)(
    3 2^{1/2} -4)
\over  2^{3/2}(2^{1/2}-1)}.
\end{array}\label{sys44aa1}
\end{equation}  
If $\theta=0$ then $h_2=0$  and $\widehat{A}(\varphi,\,0)=h_1=2^{1/2}-4$. Then, for 
 $\dsize\theta= {\pi\over 3}$ $\dsize h_1={2^{1/2}-10\over 4}$, $\dsize h_2=-{3^{1/2}\over 4}(2^{1/2}-1)$, 
 $\Pi $ at 
 $\dsize\varphi\to \pm {\pi\over 2}$ is singular and instability appears.

Summarizing results for the steady and oscillatory dynamos we have the following predictions for stability of 
field  $(\widehat{A},\, \widehat{B})$. For ${\cal D}<0$  $(A,\,B)$ is steady and   $(\widehat{A},\, \widehat{B})$ is unstable when  $\dsize\Big{|}{{\cal D}\over {\cal D}^{\rm cr}_{+} B^2}\Big{|} \gg 1$.

When $(A,\,B)$ oscillates then $(\widehat{A},\, \widehat{B})$ continues to oscillate with $(A,\,B)$ increasing the phase shift between $(\widehat{A},\, \widehat{B})$ and $(A,\,B)$. Then instability caused by the parameter resonanse may arise.
 \section{Conclusions}
 
  Here we argue, that stability of the kinematic $\alpha\omega$-dynamo problem with the  $\alpha$-effect  taken from the  the weakly-nonlinear regime near the threshold of generation can be predicted from the knowledge on the threshold of generation of the linear problem with the opposite sign of the dynamo number. It appears, that
 in spite of the fact, that the  magnetic field already saturated $\alpha$, it still can generate magnetic field  if spectra of linear problem are similar for dynamo number $\cal D$ with the opposite sign. 
      So, as $\cal D$ depends on the product of the $\alpha$ and $\omega$ effects similar analysis can be performed with  the $\omega$-quenching, usually used in geodynamo models, see e.g. \citep{Soward1978}, as well as with the feed back of the magnetic field on diffusion. It is likely, that for the more complex systems, velocity field, taken from the saturated regime,  with many exited modes will always generate magnetic field if the Lorentz force would be omitted.

So as nonlinearity (\ref{non})   has quite a general form, we consider applications of these results to some  other  dynamo models. 

Linear analysis of the axi-symmetrical $\alpha\omega$-equations gives the following, see \citep{Moffatt} and references therein:
for positive ${\cal D}$ (which is believed to be in the Earth) in presence of the meridional velocity $U_p$ the first exciting mode is dipole with $\Im\gamma=0$. Reducing of $U_p$ firstly leads to
 oscillatory dipole solution (regime of the frequent reversals \citep{B1964}). The further reduce of $U_p$ gives 
 the quadrupole oscillatory regime with larger value of ${\cal D}^{\rm cr}$. For negative  $\cal D$  and  $U_p\ne 0$
 the first mode is quadrupole with  $\Im\gamma=0$.  $U_p\to 0$ gives non-oscillatory dipole mode with decreased 
  ${\cal D}^{\rm cr}$, see for more details  \citep{Meunier1997}. In contrast to the dynamo in the disk the thresholds of generation for positive and negative 
  $\cal D$ in the sphere are of the same order and situation with stability of the field $\widehat{\bf B}$   is uncertain, and can depend on the particular form of the 
  $\alpha$- and $\omega$-effects.  Anyway, stability of $\widehat{\bf B}$ for the steady regime is more expected.

In accordance with   \citep{CT2009} shell models of turbulence demonstrate exponential growth of the magnetic field. This case, as well as 3D simulations of the turbulence in the box, which have the same instabilities,  correspond to the oscillatory regimes and using our predictions should be unstable.

In the case of  the 3D dynamo in the sphere simulations demonstrate different behaviour of $\widehat{\bf B}$ 
\citep{T2008, SSCH2009}. For  small Rayleigh numbers, when the preferred solutions is dipole (in oscillations)
  and close to the single mode structure (Case 1 in \citep{Ch2001}) $\widehat{\bf B}$  is finite. 
 Increase of the Rossby number \citep{SSCH2009} leads  to the turbulent state 
    and $\widehat{\bf B}$ becomes unstable. 

Author is grateful to A.Brandenburg and D.Sokoloff for discussions.

\end{document}